\documentclass[preprint,11pt]{JHEP3} 

\usepackage{epsfig}

\title{Deep inelastic scattering from gauge string duality in D3-D7 brane model}

\author{C.  A.  Ballon Bayona, Henrique Boschi-Filho and Nelson R. F. Braga \\
Instituto de F\'{\i}sica,
Universidade Federal do Rio de Janeiro, Caixa Postal 68528, RJ
21941-972 -- Brazil\\
E-mails: \email{ballon@if.ufrj.br}, \email{boschi@if.ufrj.br}, \email{braga@if.ufrj.br}}

\preprint{arXiv:0807.1917 [hep-th]}

\abstract{
We calculate deep inelastic structure functions 
for mesons in the D3-D7 brane model, that incorporates flavour to the AdS/CFT correspondence.
We consider two different prescriptions for the hadronic current dual:
a gauge field in the AdS bulk and a gauge field on the D7 brane.
We also calculate elastic form factors in both cases. We compare our results with other holographic models.
  }


\begin{document}

\section{ Introduction }
The AdS/CFT correspondence\cite{Maldacena:1997re,Gubser:1998bc,Witten:1998qj} relates 
string theory to superconformal gauge theories at large 't Hooft coupling.
In particular string theory in $AdS_5\times S^5 $ corresponds to a gauge theory
on the four dimensional boundary.
This exact correspondence inspired many interesting phenomenological models known as 
AdS/QCD that approximately describe important aspects of hadronic physics.

One of the simplest models, now called hard wall model, consists in breaking conformal invariance by introducing a hard cut off 
in the AdS space. The position of the cut off represents an infrared mass scale for the gauge theory.
This model was very successful in reproducing the scaling of hadronic scattering amplitudes at fixed angles
from strings in AdS space\cite{Polchinski:2001tt}.
This result was also obtained from this model using a map between bulk and boundary states in \cite{BoschiFilho:2002zs}. String theory predictions in flat space\cite{Green:1987sp} were in contrast to the experimentally observed behavior, obtained from QCD a long time ago\cite{QCD1,BRO}.
The AdS warp factor is crucial to find the correct hard scattering behavior. 
The hard wall model was also useful to estimate hadronic masses 
\cite{BoschiFilho:2002ta,BoschiFilho:2002vd,deTeramond:2005su,Erlich:2005qh,DaRold:2005zs,BoschiFilho:2005yh}. 

Another interesting model inspired in AdS/CFT is the soft wall,  that consists of a background including  AdS space and a dilaton field. 
This field acts effectively as a smooth  infrared cut off and leads to linear Regge trajectories
for mesons\cite{Karch:2006pv}.  This model was also used to calculate
masses for glueballs \cite{Colangelo:2007pt} and light scalar mesons \cite{Colangelo:2008us}. 

In the AdS/CFT correspondence for the case of $AdS_5\times S^5 $ space, the fields show up as excitations of
open strings attached to $N_c$  D3 branes. So, they are in the adjoint representation of the $SU(N_c)$ group.
In order to introduce fields that are in the fundamental representation, like quarks, one can 
add $N_f$ D7 brane probes in the space\cite{Karch:2002sh,Kruczenski:2003be,Hong:2003jm}.
In this D3-D7 branes model, open strings with an endpoint  on a D3 and the other on a D7 brane represent quarks with color and flavour. 
On the other side, mesons are described by strings with both endpoints on D7 branes. So, they correspond
to D7 brane fluctuations in the AdS background. Masses for mesons in this model were calculated in 
\cite{Kruczenski:2003be}. For an excellent review see: \cite{Erdmenger:2007cm}.

The internal structure of hadrons can be probed by various interaction processes. A very important 
one is deep inelastic scattering (DIS). This process was investigated using gauge/string duality 
by Polchinski and Strassler\cite{Polchinski:2002jw} in the context of the hard wall model.
Recently a similar investigation using the soft wall model was made in \cite{BallonBayona:2007qr}.
The hadronic structure functions found from the hard and soft wall models are different, although
they coincide at leading order. 
It is interesting to remark that hadronic elastic form factors obtained from these  
models\cite{Grigoryan:2007vg,Grigoryan:2007my,Brodsky:2007hb,Kwee:2007nq} are not the same. Other aspects 
of DIS structure functions and related hadronic processes from  AdS/QCD have been recently discussed in for example \cite{BallonBayona:2007rs,Cornalba:2008sp,Hatta:2008tn,Pire:2008zf,Albacete:2008ze,Hatta:2008qx}.

The hard and soft wall model do not take into account the flavour degrees of freedom. 
Here we will consider deep inelastic scattering using the D3-D7 brane model, where the flavour of 
the quarks is included. Elastic form factors in this model were calculated in 
\cite{Hong:2003jm,RodriguezGomez:2008zp}. 
 
This work is organized as follows: in section {\bf 2} we review deep inelastic scattering
in the hard and soft wall models.  
In section {\bf 3} we briefly describe the D3-D7 model and calculate 
the structure functions for scalar mesons considering two 
cases: the gauge field living in the entire AdS bulk and the gauge field constrained to the D7 brane. 
We present in section {\bf 4} several plots comparing the structure functions of the four models considered 
in this article. 
In section {\bf 5} we calculate the elastic form factors for both cases
and in section {\bf 6} we present our conclusions. 


\section{ Deep Inelastic Scattering in the hard and soft wall models }

In deep inelastic scattering (DIS) a lepton scatters from a hadron of momentum $P^\mu $ through the exchange of a virtual photon of momentum $q^\mu$. The final hadronic state $X$ with momentum $P_X^\mu$ is not observed. 
Then cross section involves a sum over all possible $X$.     
The DIS parameters are $q^2$ and the Bjorken parameter $x \equiv  -q^2 /2P\cdot q \,$.

The DIS cross section is calculated from the hadronic tensor  
\begin{equation}
W^{\mu\nu} \, = i \, \int d^4y\, e^{iq\cdot y} \langle P, {\cal Q} \vert \, \Big[ J^\mu (y) , J^\nu (0) \Big] 
\, \vert P, {\cal Q} \rangle \,,
 \label{HadronicTensor}
\end{equation}

\noindent where $ J^\mu(y)$ is the electromagnetic hadron current and $ {\cal Q} $ is the electric charge of the initial  hadron. The structure functions $F_1 (x,q^2) $ and $F_2 (x,q^2) $  are defined by the tensor decomposition for the spinless case\cite{Manohar:1992tz}
\begin{equation}
W^{\mu\nu} \, = \, F_1 (x,q^2)  \Big( \eta^{\mu\nu} \,-\, \frac{q^\mu q^\nu}{q^2} \, \Big) 
\,+\,\frac{2x}{q^2} F_2 (x,q^2)  \Big( P^\mu \,+ \, \frac{q^\mu}{2x} \, \Big) 
\Big( P^\nu \,+ \, \frac{q^\nu}{2x} \, \Big)
\, ,  \label{Structure}
\end{equation}

\noindent where we use the convention $\eta_{\mu\nu}={\rm diag}(-,+,+,+)$. 

On the other hand, the DIS cross section is related, by the optical theorem,  to
the forward hadron-photon Compton scattering amplitude, determined  by  the tensor 
\begin{equation}
T^{\mu\nu} \, = i \, \int d^4y e^{iq\cdot y} \langle P, {\cal Q} \vert \, {\cal T} \Big(  J^\mu (y) J^\nu (0) \Big)  
\, \vert P, {\cal Q} \rangle\,.
 \label{forwardamplitude}
\end{equation}

\noindent This tensor can be decomposed in the same way as equation (\ref{Structure}) but with 
structure functions ${\tilde F}_1 (x,q^2)  $ and ${\tilde F}_2 (x,q^2) $ 
which are related to the DIS structure functions by
\begin{equation}
\label{optical}
F_{1,2} (x,q^2) \equiv 2 \pi \,{\rm Im }\,{\tilde F}_{1,2} (x,q^2)\,.
\end{equation}

The imaginary part of the forward Compton scattering amplitude can be expressed in terms of a sum over the intermediate states $X$ with mass $M_X\,$ 
\begin{equation}
\label{Imag}
{\rm Im} T^{\mu\nu} \, = \,  2 \pi^2 \, \sum_X \,\delta \Big( M_X^2 + (P+q)^2 \, \Big) 
\langle P, {\cal Q} \vert J^\nu ( 0 )   \vert P + q,\, X \rangle\,
\langle P + q , \, X \vert J^\mu ( 0 )   \vert P, {\cal Q} \rangle\,\,.
\end{equation}


\subsection{ DIS in the hard wall model}

The hard wall model consists of an $AdS_5 \times W \,$, space
with metric $g_{MN}$: 
\begin{equation}
\label{AdS} ds^2 \equiv g_{MN} \,dx^M dx^N \,= \, \frac{R^2}{z^2}( dz^2  +
\eta_{\mu\nu} dy^\mu dy^\nu  )\,+  \,R^2 ds_W^2\,\,\,, 
\end{equation}

\noindent with  $ 0 \le z \le 1/\Lambda \,$ , where $\Lambda\,$ is an infrared cut off  interpreted as the QCD scale and $R$ is the AdS radius defined by $R^4=4\pi\, N_c\, g_s\, {\alpha'}^2$.  $W $ is a five dimensional  compact space which in the simplest case can be identified with
$\,S^5\,$. 
The prescription relating the matrix elements of a scalar hadronic $U(1)$  current to a ten dimensional supergravity   interaction action found in  \cite{Polchinski:2002jw} is the following:
\begin{equation}
(2 \pi)^4 \, \delta^4 ( P_X - P - q ) \,\eta_\mu \,  \langle P + q, X \vert J^\mu ( 0 )  \vert P,{\cal Q} \rangle
\,=\, i {\cal Q} \, \int d^{10}x \sqrt{-g} A^{ \texttt m} \Big( \Phi_i\partial_{\texttt m} \Phi_X^\ast \,-\, 
  \Phi_X^\ast  \partial_{\texttt m} \Phi_i     \Big)\,,
\label{INTERACTION}
\end{equation}

\noindent where $ \eta_\mu $ is the virtual photon polarization, $A_{\texttt m} (x)=(A_z,A_\mu)$ is a Kaluza-Klein gauge field, $ \Phi_i $ and $ \Phi_X$ are the dilaton fields representing the initial and final scalar states. This prescription is valid only in the regime $ x >> 1/ \sqrt{gN_c} $. For smaller values of $x$ 
one should include massive string states, as discussed in \cite{Polchinski:2002jw}. 

In the hard wall model the relevant gauge field solutions are the non normalizable modes
represented in terms of the Bessel function  $ \,K_1 (qz)\,$. For the scalar states with momentum $p$ the relevant solutions are normalizable and expressed in terms of  $ J_{\Delta-2} ( p z )\,$.
The mass spectrum of the final hadronic states in the hard wall implies that

\begin{equation}
\label{nova3}
\sum_X \,\delta \Big( M_X^2 + (P+q)^2 \, \Big) 
\,=\, \frac{1}{2\pi s^{1/2} \,\Lambda\,}\,.
\end{equation}

Using these results, the scalar structure functions in the hard wall model at leading order in $ \Lambda^2 /q^2 $
 take the form\cite{Polchinski:2002jw}:
\begin{equation}
F_1 (x,q^2)\,=\,0 \,\,\,;\,\,\, F_2 (x,q^2)\,=\, \pi C_0 \, {\cal Q}^2 \left( \frac{\Lambda^2}{q^2} 
\right)^{\Delta - 1} \, x^{\Delta + 1} \, (1- x)^{\Delta - 2}\,,
\label{hwsf}
\end{equation}

\noindent where $C_0$ is a dimensionless normalization constant and  $\Delta$ is the scaling dimension of the scalar state. 


\subsection{ DIS in the soft wall model}
In the soft wall model\cite{Karch:2006pv} there is an $AdS_5$ space with a static dilaton 
background field $\varphi $ chosen as $\varphi = c z^2$. This dilaton acts as an infrared cut off. 
The constant $c$, with dimension of mass squared, is related to the QCD scale.

In  \cite{BallonBayona:2007qr} we considered a ten dimensional extension of this model. 
In this case, the prescription for the supergravity regime is 
\begin{equation}
(2 \pi)^4  \delta^4 ( P_X - P - q ) \,\eta_\mu \,  \langle P + q, X \vert J^\mu ( 0 )  \vert P,{\cal Q} \rangle\,= \, i {\cal Q} \int d^{10}x \sqrt{-g}\, e^{-\varphi} \,  A^{\texttt m} 
\Big( \Phi_i\partial_{\texttt m} \Phi_X^\ast \,-\, \Phi_X^\ast  \partial_{\texttt m} \Phi_i     \Big)\,.
\label{INTERACTION2}
\end{equation}

The soft wall solutions for the gauge and scalar field involve confluent hypergeometric functions  $\,{\cal U} (1+\frac{q^2}{4c};2; cz^2) $ and $\,{\cal M} (\frac{p^2}{4c}+\frac{\Delta}{2};\Delta-1;cz^2)$, respectively, with $p=P$ or $P_X$.  
The mass spectrum of the final hadronic states implies that in the soft wall model  
\begin{equation}
\label{nova2}
\sum_X \,\delta \Big( M_X^2 + (P+q)^2 \, \Big) 
\,=\,  \frac{1}{4c}\, .
\end{equation}

Using these results,  we found 
\begin{eqnarray}
F_1 \,=\, 0 ; \; 
F_2 \,=\, 8 \pi^3 \,\frac{{\cal Q}^2}{x} \,(\Delta - 1) \,\Gamma (\Delta ) \,
\Big[ \frac{q^2}{4c}\Big]^3 \, \frac{\Gamma ( \frac{s}{4c} + \frac{\Delta}{2} -1 ) } {\Gamma( \frac{s}{4c} -  \frac{\Delta}{2} + 1 ) } \,\,\Big[ 
\frac{ \Gamma ( \frac{q^2}{4c} +\frac{s}{4c}  - \frac{\Delta}{2}  ) }
{\Gamma( \frac{q^2}{4c} + \frac{s}{4c} +  \frac{\Delta}{2} ) }\,\Big]^2\,,
\label{F2escalar}
\end{eqnarray}

\noindent which agrees at leading order in $ 4c/q^2 $  with the hard wall structure functions 
(\ref{hwsf}) as shown in \cite{BallonBayona:2007qr}.

\section{ DIS in D3-D7 system   }     

The D3-D7 system consists of the $AdS_5\times S^5 $ space with the addition of $N_f$ coincident 
D7 brane probes.
The localization of the D7 branes can be represented by writing the $AdS_5\times S^5 $  metric in spherical coordinates :

\begin{eqnarray}
ds^2_{10} &\equiv & g_{MN} \,dx^M dx^N \,=\,  \frac{r^2}{R^2}\eta_{\mu\nu}dx^\mu dx^\nu 
+ \frac{R^2}{r^2}\Big[ dr^2 + r^2 \, [ d\tilde{\theta_1}^2 + \sin^2\tilde{\theta_1}d\tilde{\theta_2}^2 \nonumber\\ &+&  \sin^2\tilde{\theta_1}\sin^2\tilde{\theta_2}(d\theta_1^2 + 
\sin^2\theta_1 d\theta_2^2 + \sin^2\theta_1 \sin^2\theta_2 d\varphi^2 ) ] \Big]\,.
\end{eqnarray}

\noindent The radial coordinate $0 \le r < \infty \,$ is related to the coordinate $z$ used in eq. (\ref{AdS}) by: $ r = R^2/z$.
Now defining the cilyndrical coordinates 
\begin{eqnarray}
\rho \equiv r \sin \tilde{\theta_1}\sin \tilde{\theta_2} \, , \, w_5 \equiv r \sin \tilde{\theta_1}\cos \tilde{\theta_2} \, , \, 
w_6 \equiv r \cos \tilde{\theta_1} \, \, ,  
\end{eqnarray}

\noindent the metric is rewritten as 

\begin{eqnarray}
ds^2_{10} \,=\, \frac{\rho^2+w_5^2 +w_6^2}{R^2}\eta_{\mu\nu}dx^\mu dx^\nu + \frac{R^2}{\rho^2+w_5^2 +w_6^2}\Big[ dw_5^2 + dw_6^2 +d\rho^2 + \rho^2 d\Omega_3^2 \Big] \, \, ,
\end{eqnarray}

\noindent where $ d\Omega_3^2  \,=\, d\theta_1^2 + \sin^2\theta_1 d\theta_2^2 + 
\sin^2\theta_1 \sin^2\theta_2 d\varphi^2 \,$.


In these new coordinates, the localization of the D7 branes can be chosen as $w_5 = 0 $, $ w_6 = L$. The metric induced on the brane is then

\begin{equation}
\label{branemetric} ds^2_8 \,=\,G_{ab} dx^a dx^b \,=\,   \frac{\rho^2 + L^2}{R^2}  \eta_{\mu\nu} \,dx^\mu dx^\nu +  \, \frac{R^2}{\rho^2 + L^2} \big( d\rho^2  + \rho^2 d\Omega_3^2, \big) \,. 
\end{equation}

\noindent Note that $\rho^2 + L^2 = r^2$ so that  the AdS radial coordinate $r$ on the brane 
is restricted to $ L  \le r < \infty $. This corresponds to an induced infrared cut off:
$m_h = L/R^2$.

Scalar mesons in this model show up as fluctuations of the D7 branes in the transversal directions 
$w_5$, $\,w_6$. For simplicity we are going to consider just one flavour: $N_f = 1$. The effective action for these fields is\cite{Kruczenski:2003be}

\begin{equation}
S_{\phi} \,=\,- 2 \mu_7 (R\pi\alpha')^2 \, \int d^{8}x \sqrt{- G} \frac{G^{ab}}{\rho^2 + L^2}
\partial_a \phi^\ast  \partial_b \phi  \,
\label{acaobrana}
\end{equation}

\noindent where $\mu_7$ is the D7-brane tension. It is convenient to rescale from now on the scalar field as $\Phi= \sqrt{2 \mu_7} R\pi\alpha' \phi$. Then the equation of motion is

\begin{equation}
\partial_a \Big[ \frac{\sqrt{- G}}{ \rho^2 + L^2 } G^{ab} \partial_b \Phi \Big] \,=\,0\,.
\label{eqmotionbrana}
\end{equation}

The solution is written in terms of a hypergeometric function $ F(a,b;c;w)\,$

\begin{eqnarray}
\Phi_{n,\ell} &=&  C_{n,\ell} e^{i p\cdot y } {\cal Y}^\ell (\Omega)  \frac{ (\rho/L)^\ell}{[ 1 + \frac{\rho^2}{L^2}]^{n+\ell+1 }} 
F( -n-\ell-1, -n;\ell+2;-\frac{\rho^2}{L^2} ) 
\label{solucaoD3D7}
\end{eqnarray}

\noindent where $C_{n,\ell}$ is a normalization constant, $ \Omega$ represents the coordinates on $S^3$, $\ell (\ell + 2) $ is the eigenvalue of the angular laplacian, related to the conformal dimension $\Delta\,$ of the hadron by: $ \Delta = \ell + 3 \,$\cite{Kruczenski:2003be}. The parameter $n$ is defined by 
$ - p^2 \,=\, 4 m_h^2 [ (n + \ell +1)(n+ \ell + 2) ] $. 

Choosing the normalization condition

\begin{equation}
R\, \int d\rho d^3\Omega   \frac{\sqrt{- G}}{ ( \rho^2 + L^2 \,)^2 } \vert  \Phi \vert^2
\,=\,1 \,,
\label{Normalization}
\end{equation}

\noindent we find that $n$ is a non negative integer and that
\begin{equation}
C_{n,\ell}\,=\, \frac{1}{\sqrt{R}} \sqrt{\frac{2 (2n+2\ell + 3)\Gamma (n + 2\ell + 3 ) }{ \Gamma(n+1) 
\Gamma^2 (\ell + 2)} } \,.
\label{Cnl}
\end{equation}

The scalar field representing the initial state 
corresponds to $ \, n=0 \,$ and $ \, p = P \,$ with mass given by 
$\, P^2 \,= - 4 m_h^2 [ (\ell +1)(\ell + 2) ] \,$. 
For the final state we have $\, p=P_X\,$ and $\, n\, $ satisfying the relation:
\begin{eqnarray}
 4 m_h^2 [ (n + \ell +1)(n+ \ell + 2) ]  =  s = - P_X^2 
= - P^2 + q^2\left( \frac 1x -1\right) \,.\label{masses}
\end{eqnarray}

\noindent Explicitly, the initial and final hadronic solutions are respectively
\begin{eqnarray}
\Phi_{i} &=&  C_{_{0,\ell}} e^{i p\cdot y } {\cal Y}^\ell (\Omega)  
\frac{(\rho/L)^\ell}{[ 1+ \frac{\rho^2}{L^2} ]^{\ell+1 }} 
 \nonumber\\
\Phi_{X} &=&  C_{_{n,\ell}} e^{i p\cdot y } {\cal Y}^\ell (\Omega)  
\frac{(\rho/L)^\ell}{[ 1 + \frac{\rho^2}{L^2}]^{n+\ell+1 }} 
F( -n-\ell-1, -n;\ell+2;-\frac{\rho^2}{L^2} ) \,.
\label{solucaoD3D7}
\end{eqnarray}

Finally, using eq.(\ref{masses}) we can calculate the sum over the masses of the final states appearing in eq. (\ref{Imag}) for the D3-D7 model:
\begin{equation}
\label{massasD3D7}
\sum_X \,\delta \Big( M_X^2 + (P+q)^2 \, \Big) 
\,\approx \,  \frac{1}{4 m_h^2 ( 2n + 2\ell + 3)}\,=
\frac{1}{4 m_h \sqrt{ s + m_h^2}}\,.
\end{equation}

Next we consider a gauge field ("photon") that will be taken as the approximate dual of the boundary hadronic current. 
We will consider two possibilities to mimic this current: a gauge field
living in the AdS bulk and a gauge field living on the D7 brane.


\subsection{Gauge field in the AdS bulk}

The interaction of the meson with the virtual photon can be represented holographically by
the interaction of a Kaluza Klein gauge field with the D7 brane scalar field.
A perturbation of the $AdS_5 \times S^5$ metric of the form $\delta g_{m\alpha} = A_m v_\alpha $,
where $m = ( r, \mu )$ and $ \alpha = (\theta_1, \theta_2 , \phi )$
induces a perturbation of the D7 brane metric of the form 
$\delta G_{{\tilde m}\alpha} = A_{\tilde m} v_\alpha $,
where ${\tilde m} = ( \rho , \mu )$ and $A_\rho = \rho/r \, A_r$. This metric perturbation
in action (\ref {acaobrana}) leads to the following interaction term

\begin{equation}
S_{int} \,=\,  \int d^{8}x \sqrt{- G} \,
\frac{v^\alpha A^{\tilde m}}{\rho^2 + L^2} \Big(
\partial_\alpha \Phi^\ast  \partial_{\tilde m} \Phi  \,
+ \partial_{\tilde m} \Phi^\ast  \partial_{\alpha} \Phi \Big)
\,.
\label{acaoint1}
\end{equation}

\noindent Using the relation between the charge ${\cal Q}$ and the Killing vector $ v^\alpha\,\,$ : $  
 R\, v^\alpha \partial_\alpha \Phi \,=\, i {\cal Q} \Phi \,$ the matrix element of the interaction action
$ \langle \, f \, \vert \, S_{int} \, \vert \, i \, \rangle \, $
takes the form (omitting the states in the notation)

\begin{equation}
 \langle  S_{int}  \rangle \,=\, \frac{ i\,{\cal Q}}{R} \, \int d^{8}x \sqrt{- G} \,A^{\tilde m} \,j_ {\tilde m}
\,,
\label{acaoint2}
\end{equation}

\noindent where 

\begin{equation}
j_ {\tilde m} \,=\, \frac{1}{\rho^2 + L^2}\,
\Big( \Phi_i   \partial_{\tilde m} \Phi^\ast_X \,-\, 
(\partial_{\tilde m} \Phi_i )\,   \Phi^\ast_X  \Big)  \,.
\label{current}
\end{equation}

\noindent is a five dimensional conserved current. The prescription 
for calculating the matrix element of the hadronic current is 
\begin{eqnarray}
(2 \pi)^4  \delta^4 ( P_X - P - q ) \,\eta_\mu \,  \langle P + q, X \vert J^\mu ( 0 )  \vert P,{\cal Q} \rangle & & \nonumber\\
 =\, \frac{ i \,{\cal Q}}{R} \, \int d^{8}x \sqrt{- G} \,A^{\tilde m} & & \!\!\!\!\!
\frac{1}{\rho^2 + L^2}\, \Big( \Phi_i   \partial_{\tilde m} \Phi^\ast_X \,-\, \Phi^\ast_X 
\partial_{\tilde m} \Phi_i \,     \Big) 
 \,.
\label{INTERACTION3}
\end{eqnarray}

\noindent 
The solution for the gauge field in the bulk, satisfying the boundary condition 
$A_\mu (\rho \to \infty , y) = \eta_\mu e^{i q\cdot y }\,$ can be written as\cite{BallonBayona:2007rs}
\begin{eqnarray}
A_{\mu}& = & \eta_{\mu}e^{iq\cdot y}\, q \frac{R^2}{r} K_1(q \frac{R^2}{r}) \nonumber \\
A_\rho & = & - \frac{i}{q^2} \eta^{\mu\nu} q_\mu \partial_\rho A_\nu\,, 
\end{eqnarray}
 
\noindent where $\,q=\sqrt{q^2}\,$ and $\,r = \sqrt{ \rho^2 + L^2}\,$. 
The conservation of the current: $ \partial_{\tilde m} \Big[ \,\sqrt{ -G } G^{{\tilde m}{\tilde n}} 
\,j_{\tilde n} \,\Big] \,=\,0\,\,$ implies that the interaction action can be rewritten as 

\begin{equation}
 \langle  S_{int}  \rangle \,=\, \frac{i \, {\cal Q}}{R}  \, \int d^{8}x \sqrt{- G} \,G^{\mu\nu} A_\mu \,\Big( 
j_\nu - \frac{i}{q^2} q_\nu \eta^{\lambda\gamma} \partial_\lambda j_\gamma \Big) \,.
\label{acaoint3}
\end{equation}

The effective four dimensional current $ j_\nu - \frac{i}{q^2} q_\nu \eta^{\lambda\gamma} \partial_\lambda j_\gamma $ is conserved. This assures the transversality of the scattering tensor.  

The matrix element of the hadronic current then takes the form

\begin{equation}
\label{MatrixBulk}
\langle P + q, X \vert J^\mu ( 0 )  \vert P,{\cal Q} \rangle \,= \, 
\, 2 \,{\cal Q}\, R \, C_{_{0,\ell}} C_{_{n,\ell}} \, \frac{q}{m_h} \,
 \Big[ p^\mu + \frac{q^\mu}{2x} \Big] \,{\cal I}_{bulk}
\end{equation}
 
\noindent where

\begin{equation}
\label{Integral}
{\cal I}_{bulk} \,=\, \int_0^1 dv v^2 (1 - v^2)^{\ell +1} \,K_1 (\frac{qv}{m_h} ) 
F(-n-\ell -1, n +\ell + 2; \ell + 2 ; 1 - v^2 )
\end{equation}

\noindent and $ v = \frac{L}{\sqrt{\rho^2 + L^2}} $.
Using these results and equations (\ref{optical}), (\ref{Imag}) and (\ref{massasD3D7}), we find 

\begin{equation}
\label{StructureD3D7}
F_1 = 0 \,\,;\,\,F_2 = 8\pi^3 {\cal Q}^2 \, \frac{\Gamma(2\ell + 4)}{\Gamma^4 (\ell + 2)} \frac{\Gamma(n + 2\ell + 3)}{\Gamma (n + 1)} \left(\frac{q^2}{m_h^2}\right)^2 \frac{{\cal I}^2_{bulk}}{x}\,.
\end{equation}

\par

We did not find a general analytical solution to the integral ${\cal I}_{bulk}$ of eq. (\ref{Integral}). 
Nevertheless, near the elastic limit $\,x \to 1\,$ we performed an analytical approximation, 
described in the appendix, and found, for $ q^2 >> m_h^2 $  

\begin{eqnarray}
F_2 &\sim& \left(\frac{4{m_h}^2}{q^2}\right)^{\ell + 2 } (1-x)^{\ell + 1}\,=\,\,
\left(\frac{4{m_h}^2}{q^2}\right)^{\Delta-1} (1-x)^{\Delta-2}\,,
\label{F2hw}
\end{eqnarray}

\noindent which agrees with the hard and soft wall model results in this regime. 

We also performed numerical calculation of the exact integral of eq. (\ref{Integral}) in order to investigate
the dependence of the structure function $F_2 $ on $q^2 $ and $x$. 
For the case $\ell = 0$ in the range $ 0.25 < x < 1 $ with $ q^2 >> m_h^2 $  we found that the structure function $F_2$ has the approximate behaviour:

\begin{eqnarray} 
F_2 &\sim& \left(\frac{4{m_h}^2}{q^2}\right)^{\ell + 2 } (1-x)^{\ell + 1}\,x^{\ell+4}
\,=\, \left(\frac{4{m_h}^2}{q^2}\right)^{\Delta-1} (1-x)^{\Delta-2}\,x^{\Delta + 1}.
\label{F2approxbulk}
\end{eqnarray}

\noindent For  $\ell = 1$ this behaviour is observed in the range $ 0.25 < x < 0.9 $.
This results are the same found for the  hard and soft wall models at leading order.

In section {\bf 4} we  plot the structure function (\ref{StructureD3D7}) comparing it  with the other models.


\subsection{Gauge field on the D7 brane}

In reference \cite{Kruczenski:2003be}, solutions for gauge fields living on the D7 brane 
were studied. These solutions were obtained from the Dirac-Born-Infeld action for the D7 brane. They obtained various types of gauge field solutions. Here we are interested in solutions for transverse photons of the form
\begin{equation}
A_\mu = \eta_\mu e^{ i q\cdot y } f (\rho ) 
\,\,;\,\, q\cdot \eta = 0 \,\,;\,\, A_\rho = 0 \,\,;\,\, A_\alpha = 0 \,\,, 
\label{campogaugebrana}
\end{equation}

\noindent where $ f ( \rho )$ satisfies the equation
\begin{equation}
 -\,\frac{q^2}{m_h^2} \frac{1}{(\zeta^2 + 1)^2} f \,+\, \frac{1}{\zeta^3} 
\partial_\zeta \Big( \zeta^3 \partial_\zeta f \Big) \,=\, 0
 \end{equation}

\noindent with $\zeta = \rho / L \,$. This corresponds to the type II gauge field of ref \cite{Kruczenski:2003be}, with no angular dependence. 
The non normalizable solution of the above equation is 
\begin{equation}
f (w) \,=\, C_q F (-\alpha , 1+ \alpha ; 2 ; w )
\end{equation}

\noindent where 
\begin{equation}
\alpha \,=\, - \frac{1}{2} \,+\, \frac{i}{2} \sqrt{\frac{q^2}{m_h^2} \,-\,1} \,\,\,\,\,;\,\,\,\,\,\, 
w \,=\, \frac{ \rho^2 }{\rho^2 + L^2  } \,\,\,\,;\,\,\,\,\,C_q \,=\,\Gamma(2+\alpha) \,\Gamma (1- \alpha)\,. 
\label{Constantes}
\end{equation}

\noindent The constant $C_q $ is determined from the boundary condition: $ A_\mu (w \to 1 )\,=\, \eta_\mu e^{i q\cdot y }\,$. This gauge field solution is dual to the boundary flavour current.  
It is important to remark that despite the fact that $\alpha$ is a complex variable, 
this solution is real since it depends only on the combination 
$\alpha (\alpha + 1)\,= - \frac{q^2}{4 m_h^2 }$. 

The interaction between this gauge field and the scalar field that lives also on the D7 brane can be obtained by imposing gauge invariance of the scalar action of eq. (\ref{acaobrana}) by introducing covariant derivatives (and again the rescaling $\Phi= \sqrt{2 \mu_7} R\pi\alpha' \phi$\, )
\begin{equation}
S_{\Phi} \,=\,- \, \int d^{8}x \sqrt{- G} \frac{G^{ab}}{\rho^2 + L^2}
\nabla_a \Phi^\ast  \nabla_b \Phi  \,,
\label{acaobranacovariante}
\end{equation}

\noindent where $\nabla_a \,=\, \partial_a \,-\, i \frac{\cal Q}{R} A_a \,$. From this action we obtain the three point 
interaction term
\begin{equation}
\langle  S_{int}  \rangle \,=\,  i\,\frac{\cal Q}{R}\, \int d^{8}x \sqrt{- G} \,G^{\mu\nu} A_\mu  \frac{1}{\rho^2 + L^2}\, \Big( \Phi_i   \partial_{\nu} \Phi^\ast_X \,-\, \Phi^\ast_X \partial_{\nu} \Phi_i \,  \Big)  \,,
\label{acaoint3}
\end{equation}

\noindent where we have used the fact that the radial and angular components of the 
gauge field vanish (see eq. (\ref{campogaugebrana})). Following a prescription similar to  eq. (\ref{INTERACTION3}) we find

\begin{equation}
\langle P + q, X \vert J^\mu ( 0 )  \vert P,{\cal Q} \rangle \,=\, 
 \,{\cal Q}\, R  \, C_{0,\ell}\,C_{n,\ell}\,
C_q \, {\cal I}_{brane} P^\mu  \,,
\label{branematrix}
\end{equation}

\noindent where $ C_{n,\ell}\,$ and $ C_q\,$ are defined in eqs. (\ref{Cnl}) and (\ref{Constantes})
and 
\begin{equation}
{\cal I}_{brane} \,=\, \int_0^1 dw w^{\ell + 1} F (-\alpha, 1+\alpha ; 2; w) \, 
F( -n-\ell -1, n+\ell + 2; \ell + 2;w)\,.
\label{Integralbrane}
\end{equation}

From eq. (\ref{branematrix}) we see that $F_1 = 0 $. 
The structure function  $F_ 2 $ is calculated  in the appendix with the 
result

\begin{eqnarray}
\label{StructurebraneReal}
F_2 &=&  8 \pi^5 {\cal Q}^2 \, \frac{\Gamma(2\ell + 4)}{\Gamma^2 (\ell + 2)} 
\frac{\Gamma(n + 2\ell + 3)}{\Gamma (n + 1)}\frac{1}{x}\, \left( \frac{q^2}{ 4 m_h^2}\right)^3 
\,  \frac{1}{ \cosh^2\,( \pi \sqrt{ \frac{q^2}{4 m_h^2 } \,- \frac{1}{4}\,}\,\,)} \nonumber\\  
&  \times& \Big\{ \sum_{j=n}^\infty \,\frac{  
\Gamma (\ell + 2 + j) \,\Gamma (j+1) }{  \,(j+1)\, \Gamma (n+ 2 \ell + 4 + j)\, \Gamma (j+1-n) }\,
\prod_{k=1}^{j} \, \Big[ 
 \frac{ q^2 }{ 4 m_h^2 \, k^2}
\,+\, \frac{k-1}{k} \,\Big]\,\Big\}^2\,.
\end{eqnarray}

This expression can be approximated  in the regime $\, x<< 1\, $. 
In this case, as described in the appendix,  the above equation reduces to

\begin{equation}
\label{Structurebrane2}
{{F_2}\vert}_{x<<1} \approx \pi^5 {\cal Q}^2  \, \frac {(1.27)^2}{2}\, \Gamma(2\ell + 4)\, (\ell+2)^2 
\left(\frac{ 4 m_h^2  }{q^2}\right)^{\ell+2}\, x^{\ell+4} \,.
\end{equation}

Note that in this model $\Delta=\ell+3$ so that this structure function reproduces  the same dependence on $q^2$ and on $x$ as in the hard wall model of eq. (\ref{hwsf}), in this regime of $\, x<< 1\, $.

We also performed a numerical analysis of the structure function $F_2$ of eq. 
(\ref{StructurebraneReal}) for different regimes. We found that the dependence on $q^2$ at leading order, for $ 0.1 < x < 1$,  is approximately equal to the hard and soft wall models. 
In section {\bf 4} we present  plots of this brane structure function, comparing with the other models, for different regimes of $x$ and $q^2$.


\section{Comparison of the models}

Here we compare our results for the structure function $F_2$ for D3-D7 brane model with the gauge field in the bulk, eq. (\ref{StructureD3D7}), and on the brane, eq. (\ref{StructurebraneReal}), with those of the hard and soft wall models, eqs. (\ref{hwsf}) and 
(\ref{F2escalar}), respectively. Since these models have different parameters representing the infrared cutoff, it is convenient to introduce a dimensionless variable $Q$ identified as $q/2m_h$ in D3-D7 model, $q/\Lambda$ in the hard wall and $q/2\sqrt{c}$ in the soft wall model. 

For $Q=75$ we show in Figure 1 the dependence of the structure function $F_2$ with the Bjorken parameter $x$, in the range $0.25 \le x < 1$, for $\ell=0,1,2$ and 3. Each plot contains the curves for the four models. Note that the D3-D7 bulk structure function coincides with those of the hard and soft wall models, represented by the blue lines. The D3-D7 model with gauge field on the brane is represented in red lines. 
Note that the maximum of these functions occur in different values of $x$, indicating a different dependence
on this variable.  
All the structure functions are normalized setting their maximum values to one in each plot. 
For other values of $Q$, we found structure functions behaviour similar to that exhibited in Figure 1, 
as long as  $Q >> 1 $. 

\FIGURE{\epsfig{file=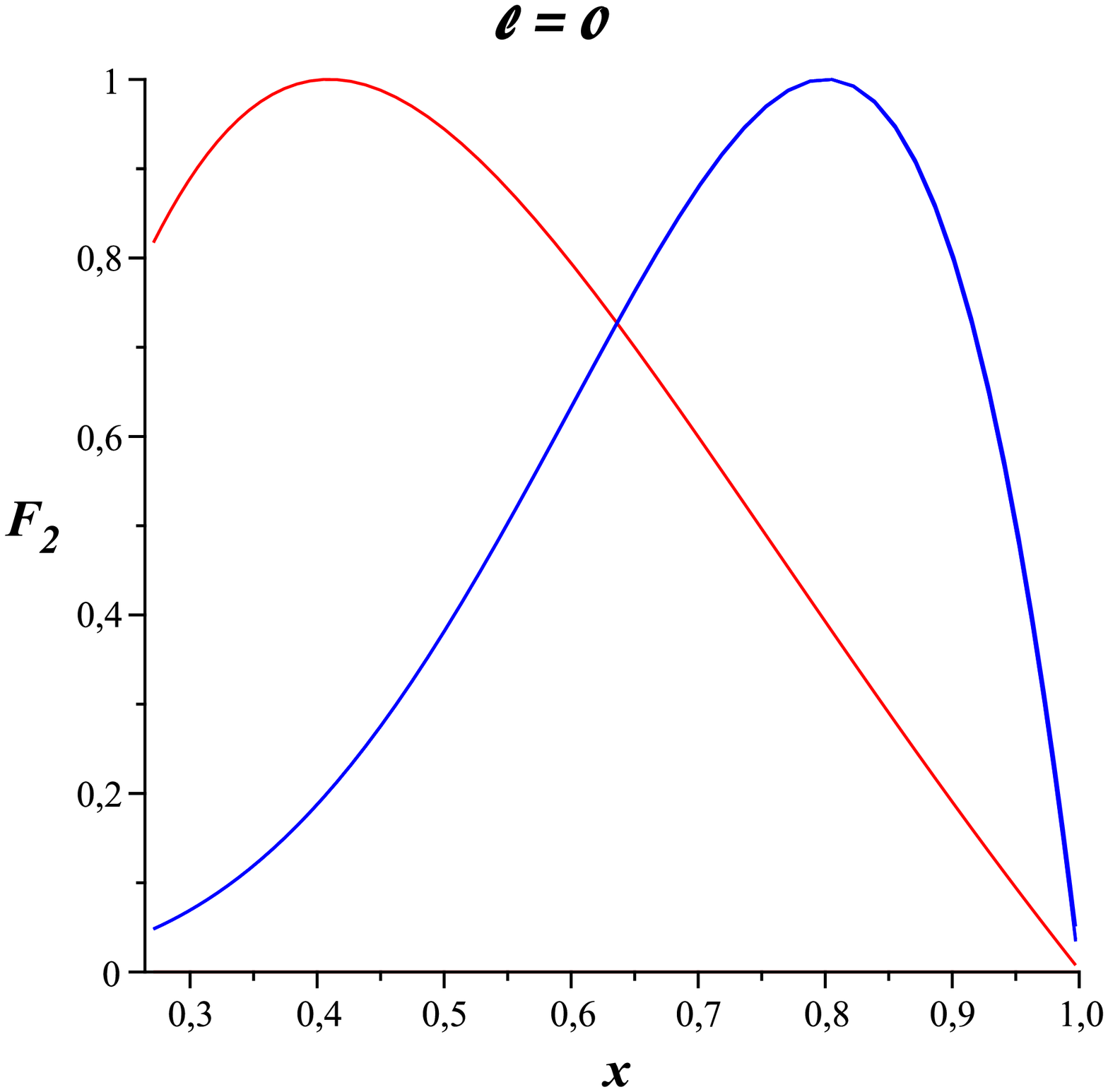,width=5cm}\epsfig{file=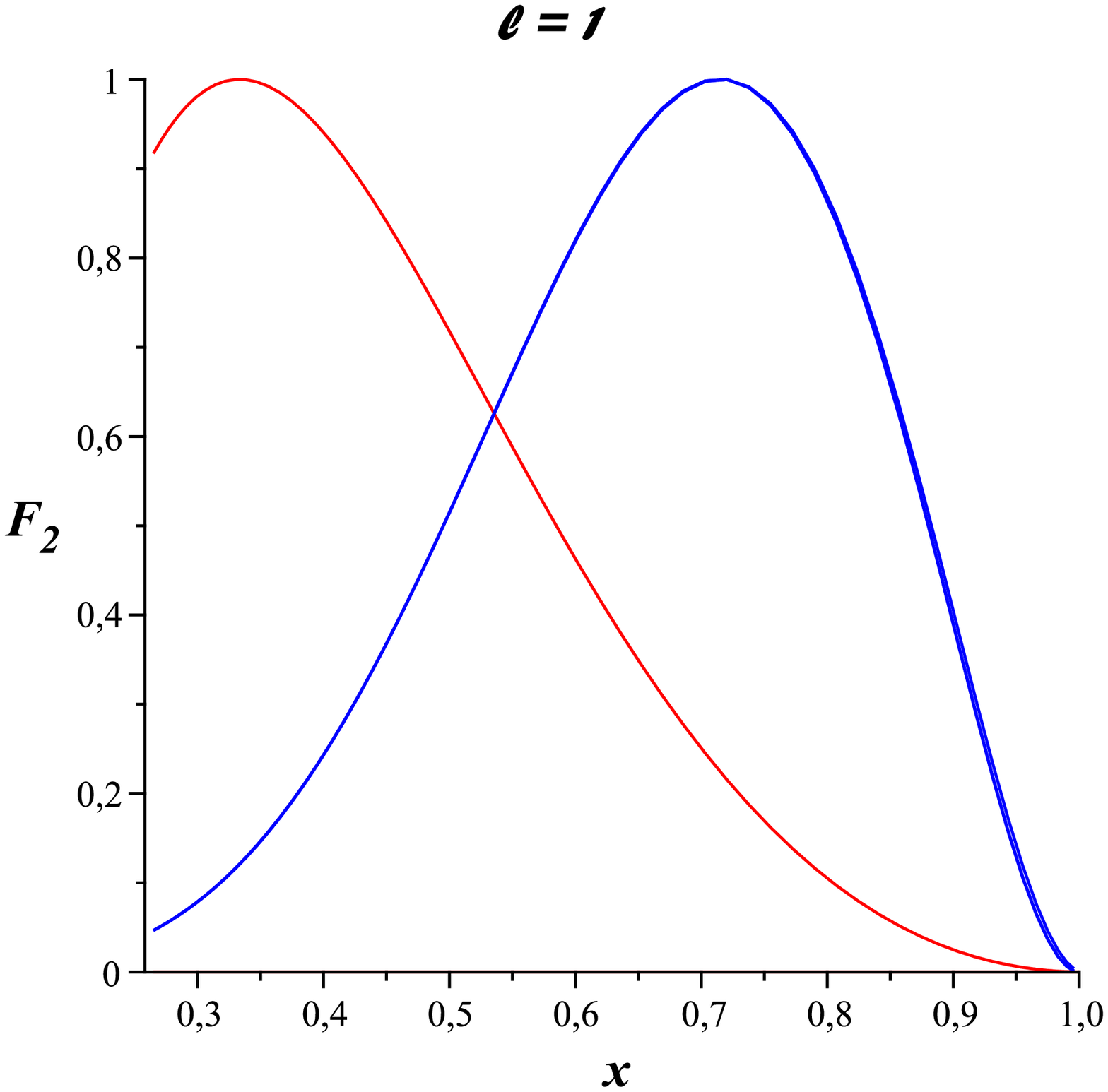,width=5cm}
\epsfig{file=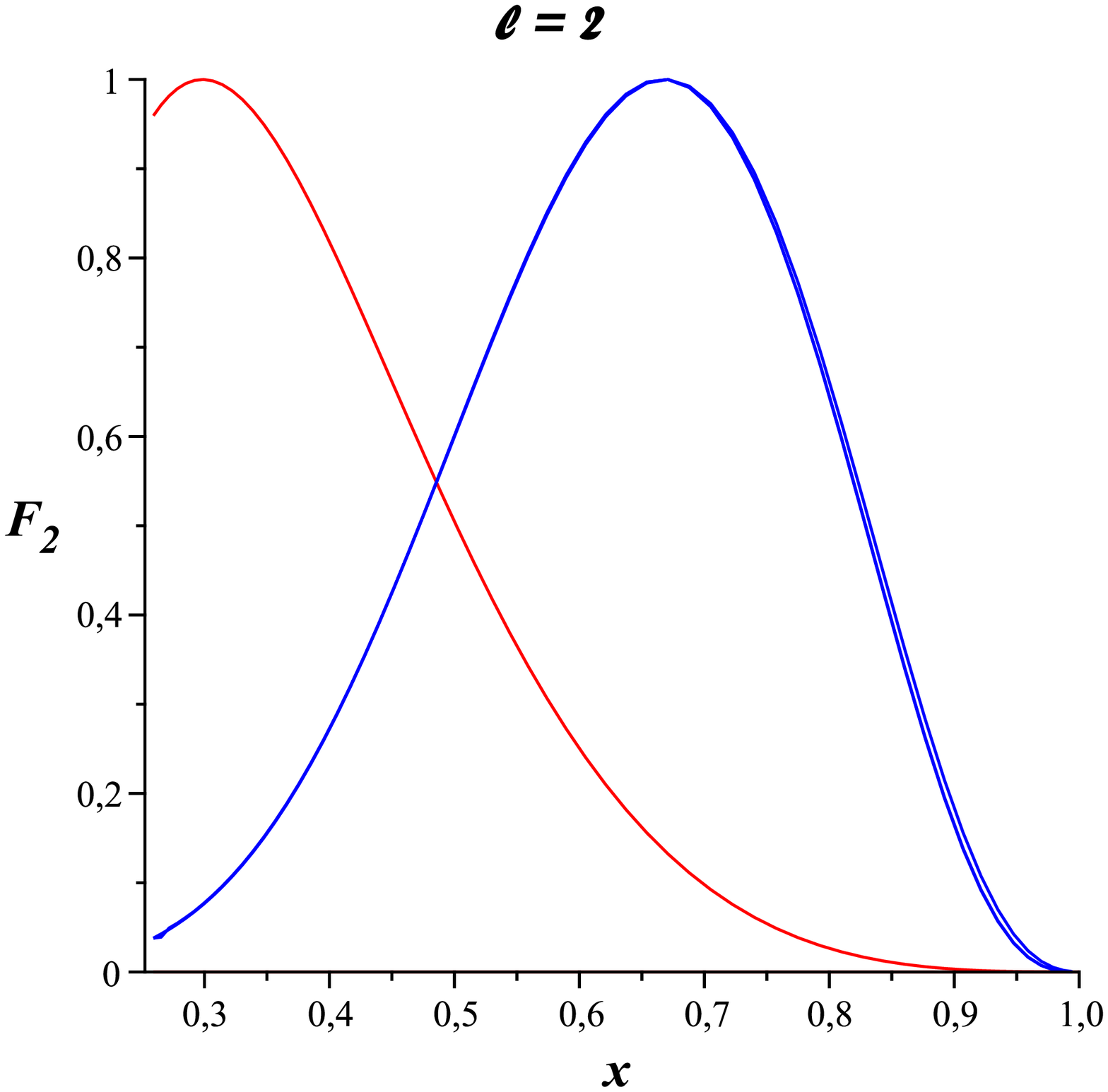,width=5cm}\epsfig{file=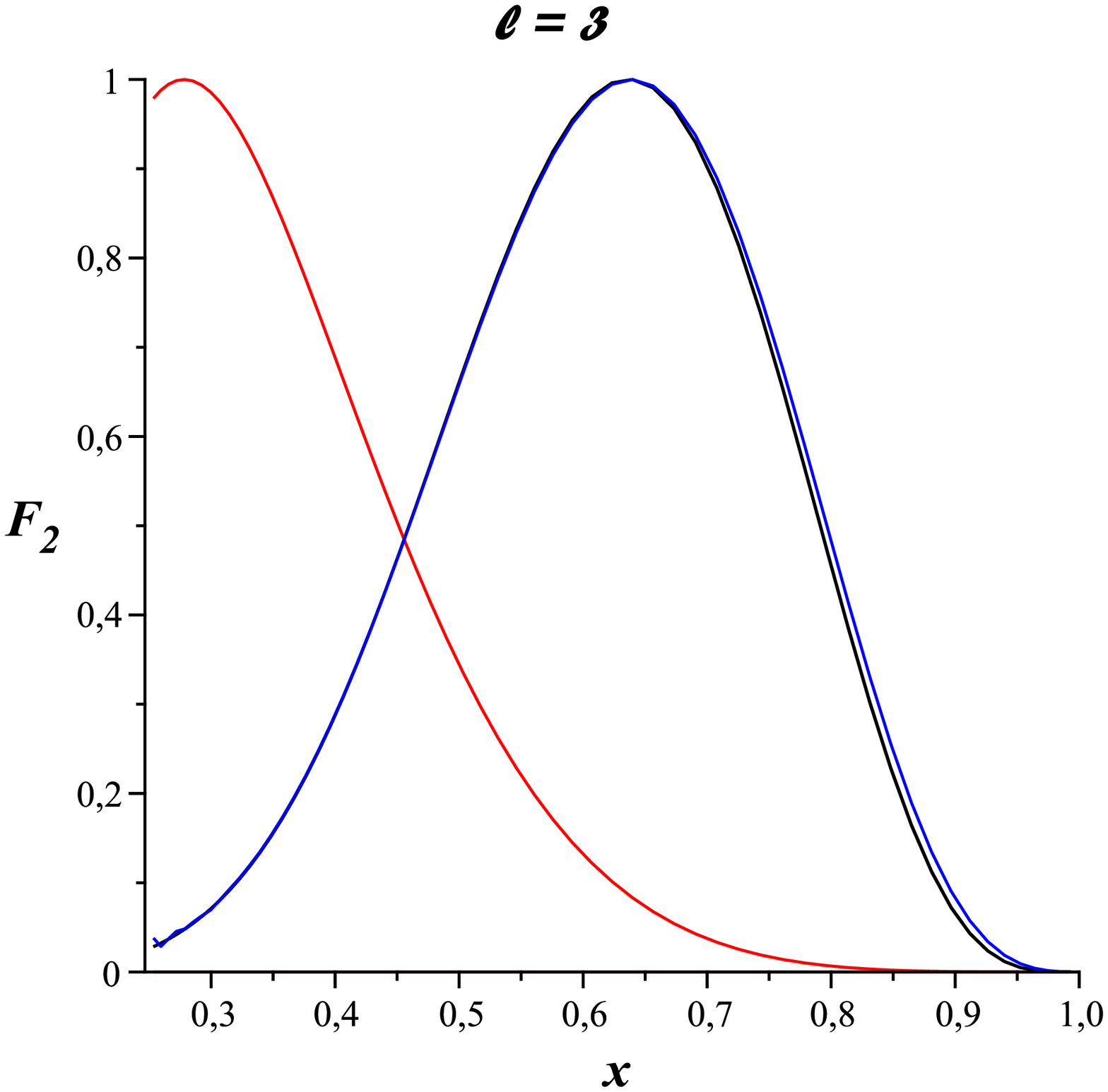,width=5cm}
\caption{The structure functions $F_2$ versus the Bjorken variable $x$ for $\ell=0,1,2,3$ and $Q=75$. The red lines represent the brane results. The blue lines represent the other models.}\label{F2_Q75_l0123}}

In Figure 2 we illustrate the small $x$ behavior of the structure functions. The bulk structure function has a singular behavior which is shown in the first plot. In the second plot, we compare $F_2$ for the brane (red line) with hard and soft wall (black line). The normalization constants are the same used in Figure 1.
These plots were obtained using $Q = 75$ but we found the same behaviour for other values of $Q >> 1$. 

\FIGURE{\epsfig{file=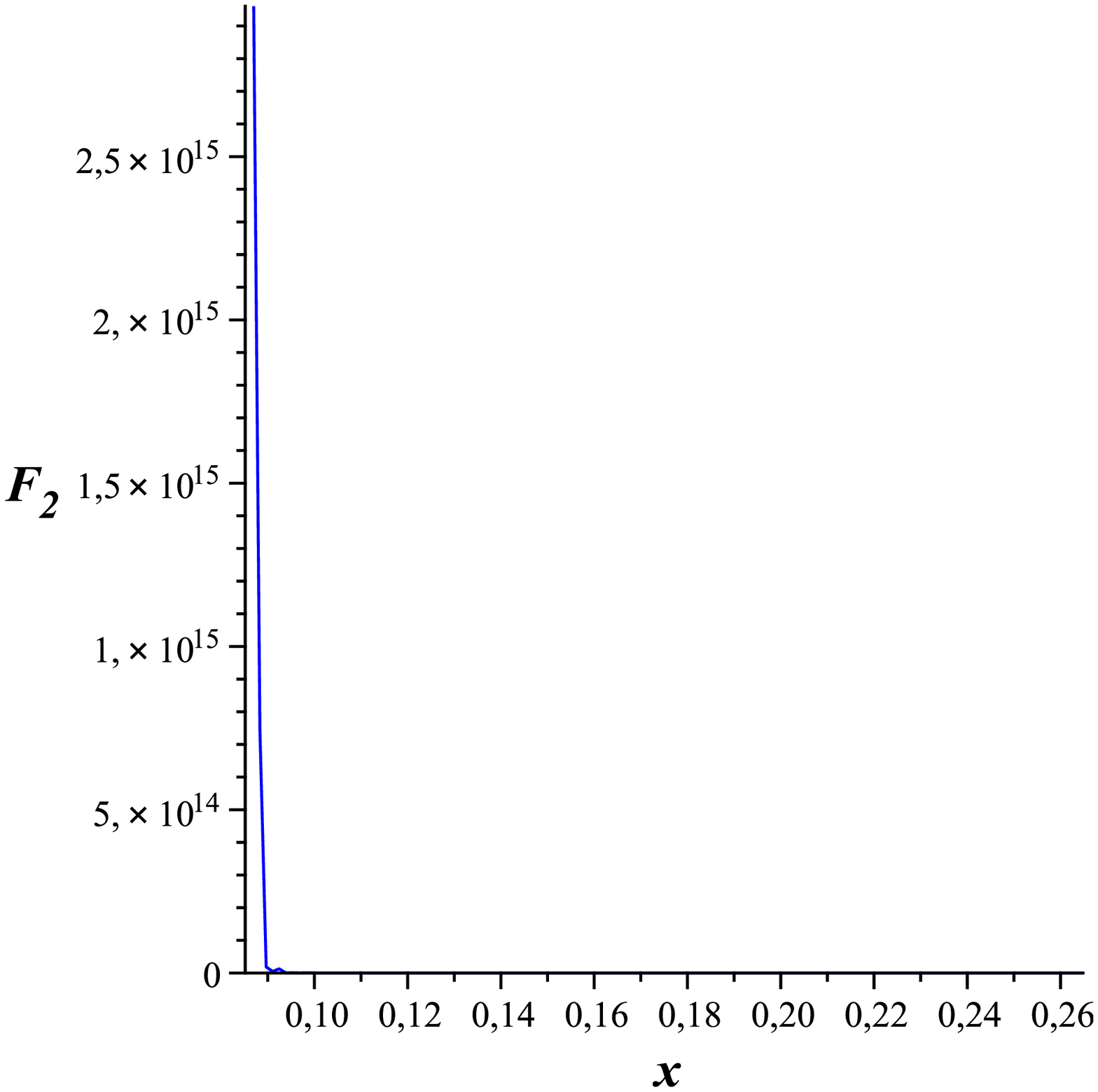,width=5cm}
\epsfig{file=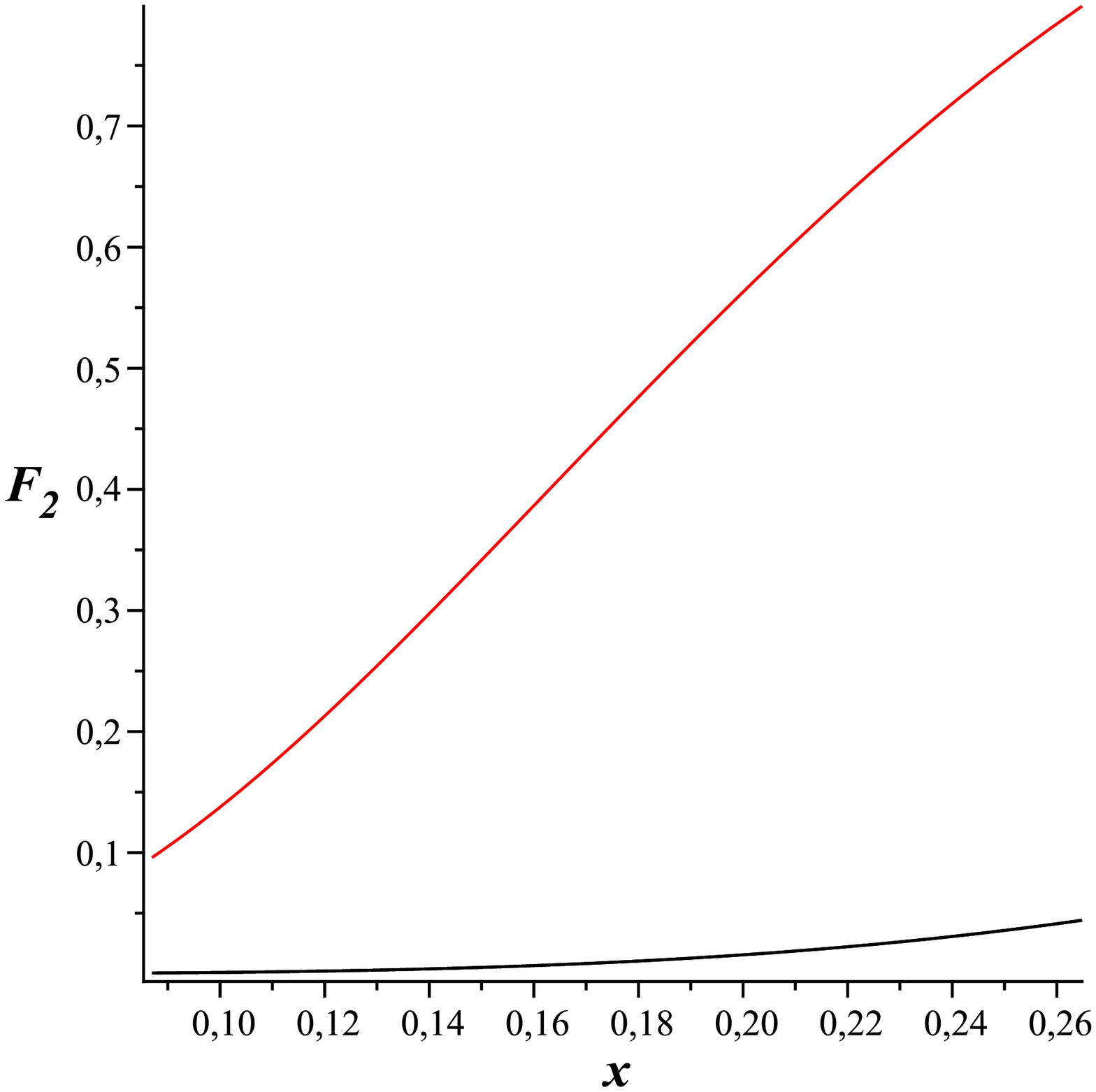,width=5cm}
\caption{The structure functions $F_2$ at small $x$ for $\ell=0$ and $Q=75$. The first plot corresponds to the bulk case (blue). In the second plot the red line represents the brane structure function, while the black line corresponds to the hard and soft wall models.}
\label{F2_Q75_small_x}}

We also studied the dependence of $F_2$ on $Q^2$. In Figure 3 we plot $-\ln(F_2(Q)/F_2(Q_0))$ as a function of $\ln(Q^2/Q_0^2)$, for various values of $x$ and $\ell $.  In the first plot $x=0.5$, $Q_0= 45 $, and $\ell=0,1,2$. In this case the structure functions for the four models coincide. Each line in this plot represents a different value of $\ell$. This plot shows that the structure functions behaves as $Q^{-2(\ell+2)}$, for this value of $x$. 
The second plot corresponds to $x=0.9$, $Q_0 = 12$  and $\ell=0$. In this case the brane is represented by the red line while the others are represented by the blue line. 
This indicates that the brane structure function decreases slightly faster than the others, that behave as $Q^{-4}$. For small values of $x$ the behavior of the bulk structure function is not well described by a power of $Q^2$ if $Q >> 1$. This is evident from the highest values of $Q$ in the  third plot of Fig. 3, where $x=0.25$ and $Q_0 = 39$.

\FIGURE{\epsfig{file=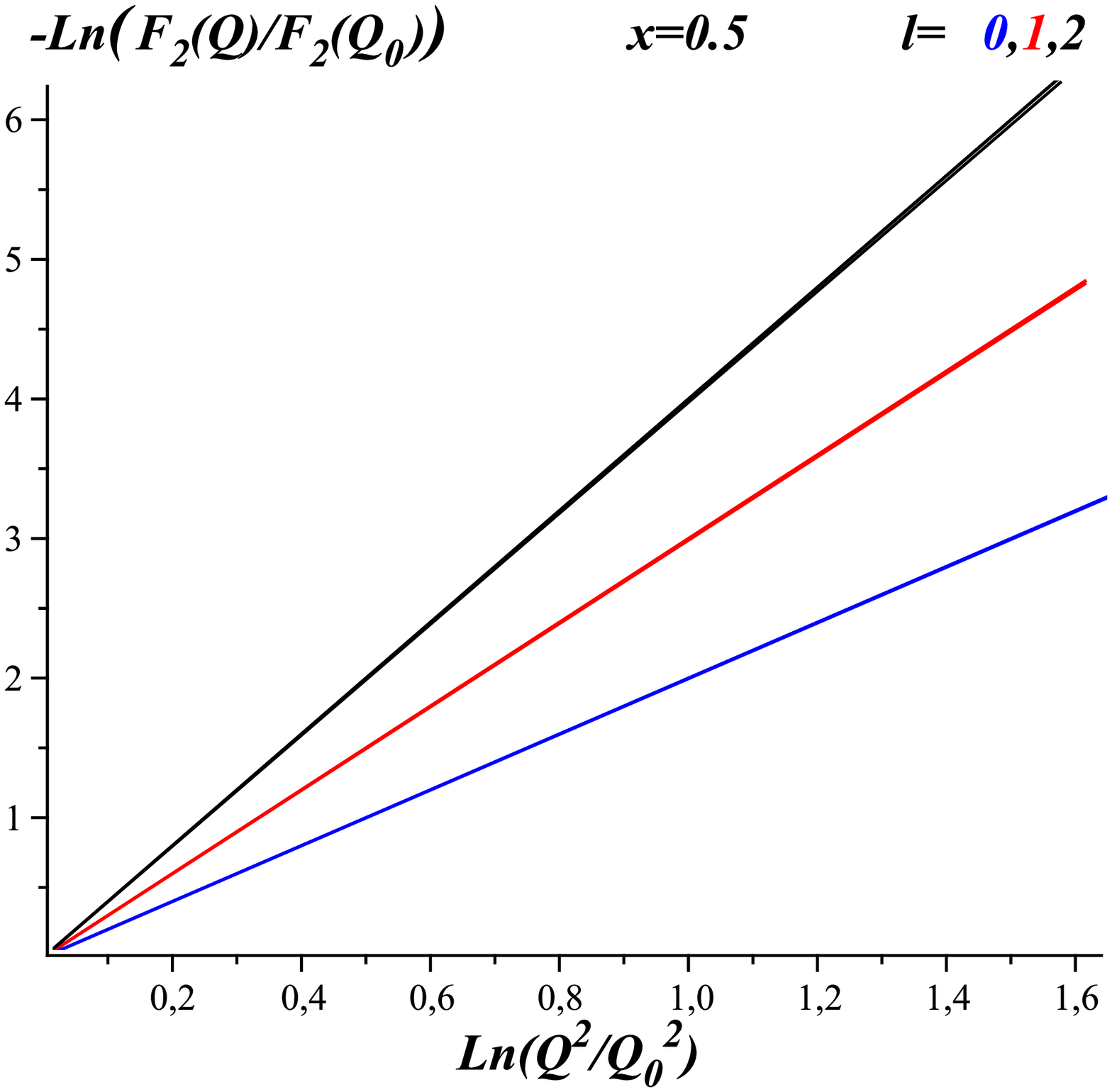,width=6cm}
\epsfig{file=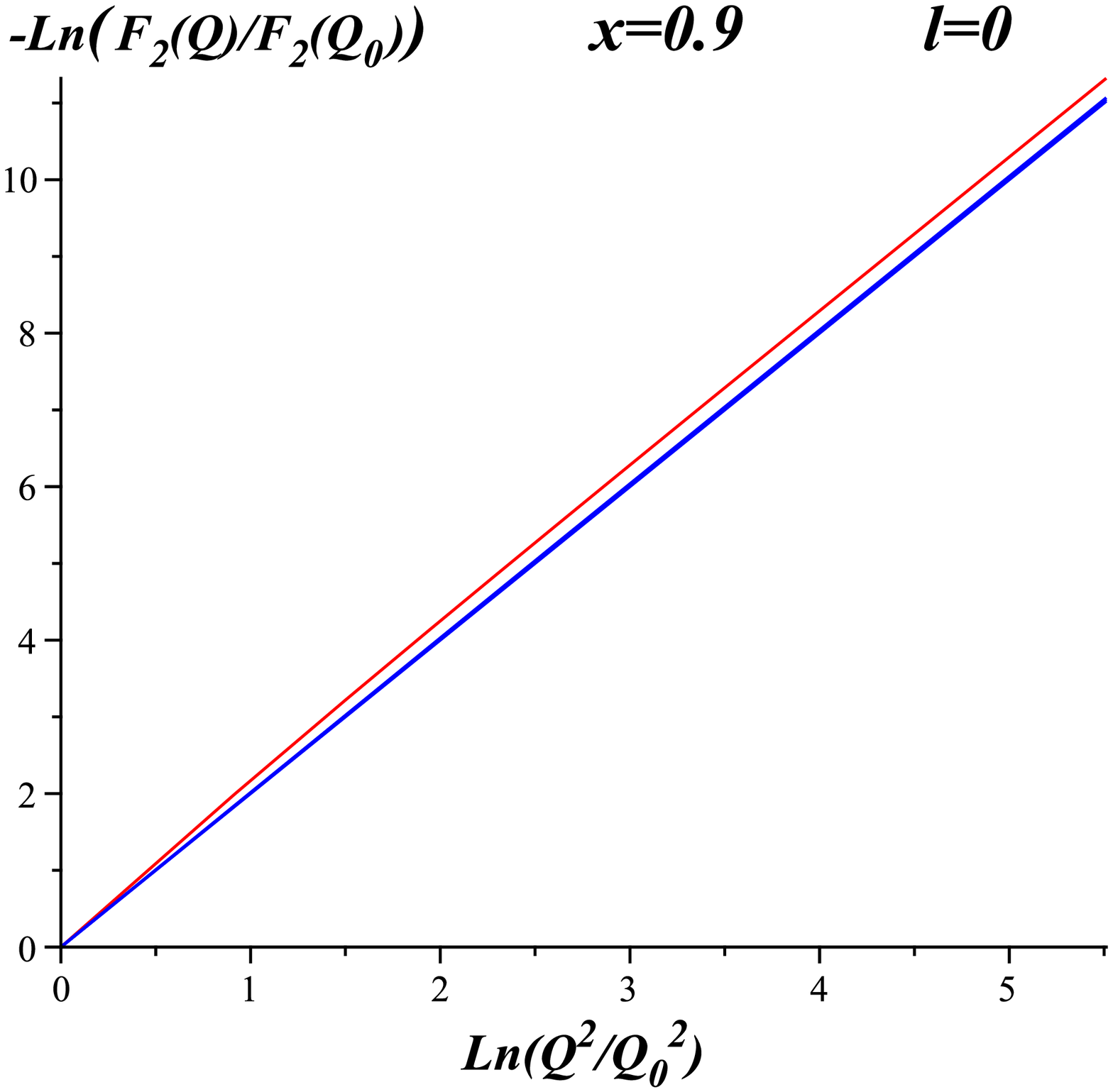,width=6cm}
\epsfig{file=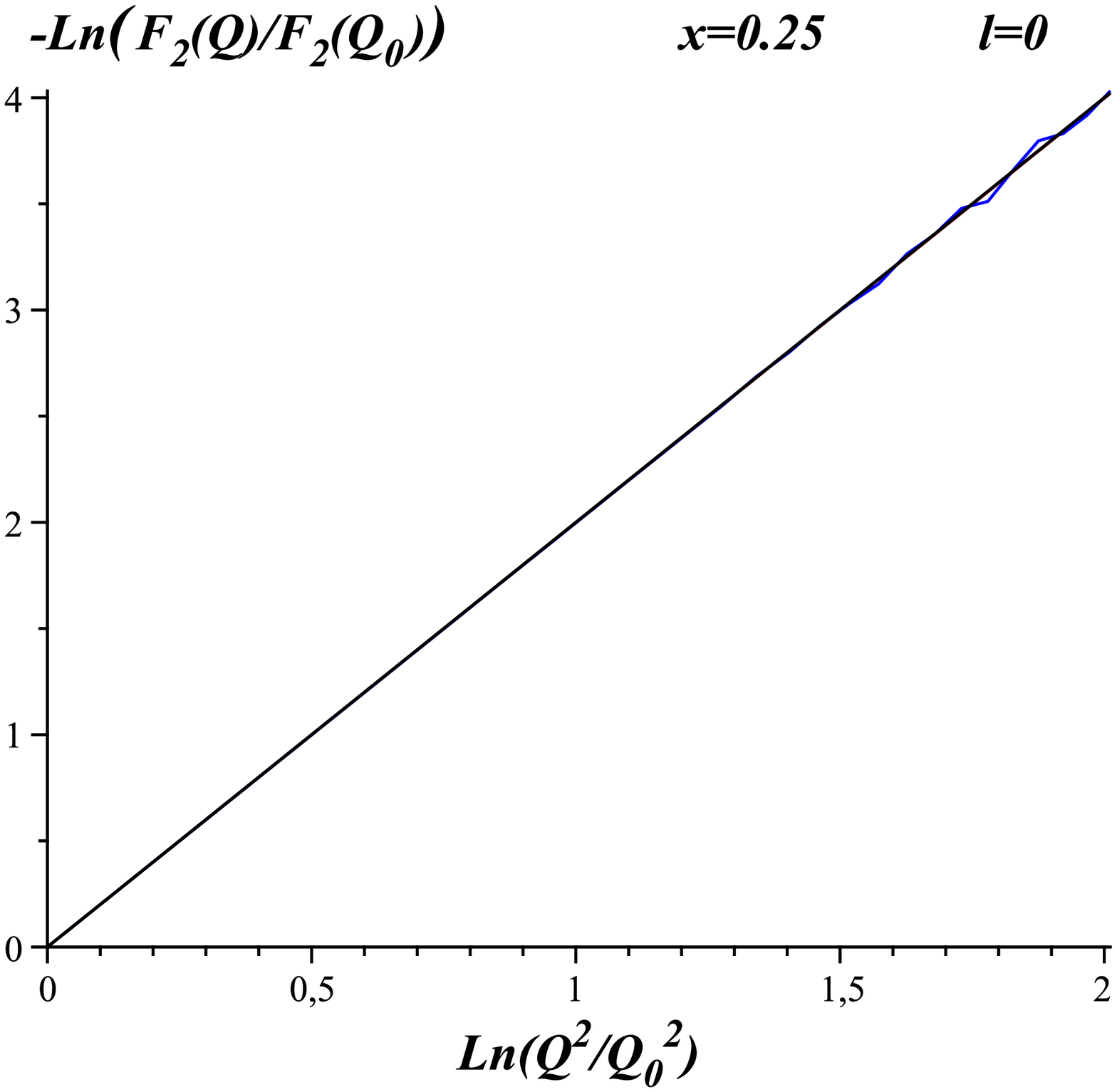,width=6cm}
\caption{The $Q^2$ dependence of the structure functions $F_2$ for the four models considered.
In the first plot, with  $x=0.5$, $Q_0= 45 $, we have $\ell=0$ ((blue line), $\ell = 1$ (red line) and
$\ell = 2$ (black line). Each line in this plot represents the coinciding results for the four models.
In the second plot $x = 0.9$ $ \ell = 0$ and the red line represents the brane case while the blue line 
represents the other three models. In the third plot, with $x = 0.25$ and $\ell = 0 $, all the structure functions coincide except for the bulk case that oscillates for higher values of $Q$. 
} \label{F2_Q2}}


\section{ Elastic form factors}

Hadronic elastic form factors can also be calculated from the results of the previous section
for the gauge field in the bulk and on the brane.
These form factors are defined in terms of the matrix element of the hadronic current
with the condition $n=0$ which corresponds to the elastic regime $x=1$. Explicitely, the form factor $F(q^2 )$ 
is defined by
\begin{equation}
\langle P + q, X \vert J^\mu ( 0 )  \vert P,{\cal Q} \rangle \,= \,2\, F(q^2)\,   [2P^{\mu}+q^{\mu}] \,.
\end{equation}  

For the case of the gauge field in the bulk, using eqs. (\ref{integral5}) and (\ref{MatrixBulk})
we find
\begin{eqnarray}
\label{formfactor}
F(q^2)&=& \frac{Q}{2} \,  \Gamma(2 \ell + 4)(\ell+2)\Big(\, \frac{4m_h^2}{q^2}\Big)^{\ell+2} \nonumber \\
&=& \frac{Q}{2}\, \Gamma(2\Delta-2)(\Delta-1)\Big(\, \frac{4m_h^2}{q^2}\Big)^{\Delta-1} \, .
\end{eqnarray}

Considering now the gauge field on the brane we can calculate the corresponding elastic form 
factor. As show in the appendix, the result is exactly the same  obtained for the gauge field in the bulk. 
This result has been obtained before in \cite{Hong:2003jm} for the D3-D7 model, using  different 
scalar field solutions.


It is interesting to note that these results are also in agreement (at leading order in $q^2$) with those coming from the hard and soft wall models \cite{Grigoryan:2007vg,Grigoryan:2007my,
Brodsky:2007hb}.

\section{Conclusions}

We calculated hadronic deep inelastic structure functions and elastic form factors using the D3-D7 brane model.
This model incorporates flavour to  AdS/CFT, based on a top-down approach. We have considered two possibilities for the gauge field dual to the hadronic current: a Kaluza Klein gauge field that lives in the AdS bulk and a D7 world-volume gauge field. They correspond to different boundary current operators. 
In the first case the boundary current is associated with the supersymmetry group SO(4) while in the second case with the flavour symmetry group $U(N_f) $. 

For intermediate values of $x$, the structure functions of the four models considered (gauge field in the bulk, gauge field on the brane, hard and soft wall) present for $\ell = 0, 1, 2$ approximately the same dependence on $q^2$:  $ (q^2)^{1-\Delta}\,$.  
For small values of $x$  the bulk structure function has an oscillating behaviour while the other 
models preserve the same power law dependence. For large values of $x$ and $\ell =0 $ we found a small deviation of the brane structure function from  $ (q^2)^{1-\Delta}\,$.

The dependence on the Bjorken variable $x$ was illustrated in figs. 1 and 2.
We found that for the gauge field in the bulk the structure function behaves as 
$ (1 - x)^{\Delta - 2} x^{\Delta +1} \,$ for intermediate values of $x$, coinciding with the hard and soft wall models. 
For the gauge field on the brane we find a different behaviour.  In particular, the localization 
of the maximum value of the structure function differs from the other three models.
This fact may be related to the different interpretation of the boundary current operators.
For the gauge field in the bulk the dual current is related to supersymmetry, as it happens in the  hard and soft wall models. This contrasts with the case of the gauge field on the 
D7-brane where the dual current corresponds to flavour symmetry. 

We considered in this article massless D7 scalar fields. This leads to a relation between the conformal dimension $\Delta$ of the hadronic operator and the angular quantum number on $S^3$:
$\Delta = \ell + 3$.  
Another different and interesting possibility would be to consider $\Delta $ to be independent of $\ell$. 
If we follow this approach, an eight dimensional mass for the D7 scalar field would be necessary. 
In this case the supergravity approximation, used in this work,  would be valid only for the 
regime  $x >> (\sqrt{gN_c})^{-1}\,$, so that the  eight dimensional mass would be negligible with respect to the string scale $ 1/ \sqrt{\alpha' }\,$.
This approach would be analogous to that considered, for ten dimensional scalar fields,
when studying DIS in the hard and soft wall models.


\bigskip

\noindent {\bf Acknowledgments:} We thank David Mateos for  important discussions. 
The authors are partially supported by CLAF, CNPq and FAPERJ. 
N.R.F.B. thanks the Institute for Nuclear Theory at the University of Washington for kind hospitality and the U.S. Department of Energy for partial support during the completion of this work.


\appendix\section{ Calculations of the structure functions and form factors }

\subsection{Gauge field in the bulk }

Here we present an approximation for the integral ${\cal I}_{bulk}$ of eq. (\ref{Integral})  
valid for $\,x \to 1\,$ and $q^2 >> m_h^2$. Using the property 
$$F( -n-\ell-1, -n+\ell+2;\ell+2;1-v^2) = v^{2(n+\ell+1)}\,F( -n-\ell-1, -n;\ell+2;1-\frac{1}{v^2}) $$ 
and expanding the hypergeometric function one finds
\begin{eqnarray}
{\cal I}_{bulk} &=& \sum_{i=0}^{n}\frac{(-n-\ell-1)_i\;(-n)_i}{(\ell+2)_i\; i!} \; (-1)^i\; {\cal I}_i\,\,,
\label{integral2}
\end{eqnarray}
where $(a)_i $ is the Pochhammer symbol and 
\begin{eqnarray}
{\cal I}_i = \int_0^1 dv\; v^{2(n+\ell+2-i)}\;(1-v^2)^{\ell+1+i}\; K_1\left(\frac{q}{m_h}v\right)  \,.
\end{eqnarray}

\noindent Near the elastic limit we have $ q^2/m_h^2 >> n^2 $. In this case the integrand decreases rapidly, because of the Bessel function $K_1$,
and the relevant contribution to the integral 
comes from the region $0\le v \le m_h/q$. 
Then we can use the approximation $(1-v^2)^{\ell+1+i} \approx 1$ in the integrand and the integral domain can be extended to $0\le v <  \infty $ 

\begin{eqnarray}
{\cal I}_i &\approx & \int_0^\infty dv\; v^{2(n+\ell+2-i)}\; K_1\left(\frac{q}{m_h}v\right) \nonumber\\
 &=& \frac 14 \left(\frac{2m_h}{q}\right)^{2n+2\ell+5-2i}\; \Gamma(n+\ell+3-i)\Gamma(n+\ell+2-i)\,\,.
\label{integral4}
\end{eqnarray}

Substituting this result in the ${\cal I}_{bulk}$ integral and using the property   
$$
(-a)_i= (-1)^i\frac{\Gamma(a+1)}{\Gamma(a+1-i)}$$

\noindent one finds 

\begin{eqnarray}
{\cal I}_{bulk} &=& \left(\frac{2m_h}{q}\right)^{2n+2\ell+5} \frac{\Gamma(n+\ell+3)}{4} \;
\Gamma(n+\ell+2)\; 
_1F_2( -n;\ell+2,-n-\ell-2;-\frac{q^2}{4{m_h}^2})\,\,,\nonumber\\
\label{integral5}
\end{eqnarray}

\noindent where $_1F_2 (a;b,c;z) $ is another  hypergeometric function. Then, using this result in eq. (\ref{StructureD3D7}) we find an approximation for the structure function near the elastic limit

\begin{eqnarray}
\label{StructureD3D7final}
F_2 &=& 8\pi^3 {\cal Q}^2 \, \frac{\Gamma(2\ell + 4)}{\Gamma^4 (\ell + 2)} 
\frac{\Gamma(n + 2\ell + 3)}{\Gamma (n + 1)} \left( \frac{q^2}{ 4 m_h^2}\right)^{-2n-2\ell-3} \frac{{1}}{x}\,\cdot\nonumber\\ 
&\cdot& \Big( \Gamma(n+\ell+3)\;
\Gamma(n+\ell+2)\; _1F_2( -n;\ell+2,-n-\ell-2;-\frac{q^2}{4{m_h}^2}) \Big)^2\,.
\end{eqnarray}

Taking just the dominant term of the hypergeometric function we find that, for $ x\to 1$ 

\begin{eqnarray}
F_2 &\sim& \left(\frac{4{m_h}^2}{q^2}\right)^{\ell + 2 } (1-x)^{\ell + 1}\,=\,\,
\left(\frac{4{m_h}^2}{q^2}\right)^{\Delta-1} (1-x)^{\Delta-2}\,.
\end{eqnarray}

\subsection{Gauge field on the brane }

Using eqs. (\ref{branematrix}) ,(\ref{Integralbrane}) and the relation
\begin{equation}
\Gamma(2+\alpha) \,\Gamma (1- \alpha)\,= \frac{ \pi q^2 }{ 4 m_h^2 
 \cosh\,( \pi \sqrt{ \frac{q^2}{4 m_h^2 } \,- \frac{1}{4}\,}\,\,)}\,,
\end{equation}

\noindent the structure functions can be written as 
\begin{equation}
\label{Structurebrane}
F_1 = 0 \,\,;\,\,F_2 = 8 \pi^5 {\cal Q}^2 \, \frac{\Gamma(2\ell + 4)}{\Gamma^4 (\ell + 2)} \frac{\Gamma(n + 2\ell + 3)}{\Gamma (n + 1)} \left( \frac{q^2}{ 4 m_h^2}\right)^3 
\,\frac{1}{ \cosh^2\,( \pi \sqrt{ \frac{q^2}{4 m_h^2 } \,- \frac{1}{4}\,}\,\,)}
\frac{{\cal I}^2_{brane}}{x}\,.
\end{equation}

In order to evaluate ${\cal I}_{brane}$ we use the identity
\begin{equation}
F( n-\ell -1, n+\ell + 2; \ell + 2;w)
 \,=\, (1-w)^{n+\ell+1}\,
F( n-\ell -1, - n; \ell + 2; \frac{w}{w-1})
\label{Prop1}
\end{equation}

\noindent and expand both hypergeometric functions. Then, we can perform the integral and find
\begin{eqnarray}
{\cal I}_{brane} &=&  \Gamma (n+\ell + 2) \sum_{j=0}^\infty  \frac{ ( -\alpha)_j (1+\alpha)_j 
\Gamma (\ell + 2 + j)}{ (2)_j \,j! \Gamma (n+ 2 \ell + 4 + j) } 
\sum_{i=0}^n \frac{ (-n)_i (\ell +2 + j)_i }{ (\ell +2)_ i \,i! }
\nonumber\\
&=& (-1)^n \,\Gamma (\ell + 2) \, \sum_{j=n}^\infty \,\frac{ ( -\alpha)_j (1+\alpha)_j 
\Gamma (\ell + 2 + j) \,\Gamma (j+1) }{ (j!)^2 \,(j+1)\, \Gamma (n+ 2 \ell + 4 + j)\, \Gamma (j+1-n) } \,\,.
\label{Integralbrane2}
\end{eqnarray}

This sum can be expressed in terms of real variables by using the relation
\begin{equation}
 \frac{ ( -\alpha)_j (1+\alpha)_j }{ (j!\,)^2 }
\,=\, \prod_{k=1}^{j} \, \Big[ 
 \frac{ ( -\alpha) (1+\alpha) }{ k^2}
\,+\, \frac{k-1}{k} \,\Big]
\,=\, 
\prod_{k=1}^{j} \, \Big[ 
 \frac{ q^2 }{ 4 m_h^2 \, k^2}
\,+\, \frac{k-1}{k} \,\Big]\,.
\label{pochhammer3}
\end{equation}

So that, the structure function takes the form

\begin{eqnarray}
F_2 &=&  8 \pi^5 {\cal Q}^2 \, \frac{\Gamma(2\ell + 4)}{\Gamma^2 (\ell + 2)} 
\frac{\Gamma(n + 2\ell + 3)}{\Gamma (n + 1)}\frac{1}{x}\, \left( \frac{q^2}{ 4 m_h^2}\right)^3 
\,  \frac{1}{ \cosh^2\,( \pi \sqrt{ \frac{q^2}{4 m_h^2 } \,- \frac{1}{4}\,}\,\,)} \nonumber\\  
&  \times& \Big\{ \sum_{j=n}^\infty \,\frac{  
\Gamma (\ell + 2 + j) \,\Gamma (j+1) }{  \,(j+1)\, \Gamma (n+ 2 \ell + 4 + j)\, \Gamma (j+1-n) }\,
\prod_{k=1}^{j} \, \Big[ 
 \frac{ q^2 }{ 4 m_h^2 \, k^2}
\,+\, \frac{k-1}{k} \,\Big]\,\Big\}^2\,.
\end{eqnarray}

\noindent {\bf Approximation for $\, x<< 1\, $}

It is possible to approximate the sum above in the regime of $\, n^2 >> q^2/4m_h^2 \, $, which means $\, x<< 1\, $. In this case the above equation reduces to 

\begin{eqnarray}
\prod_{k=1}^{j} \, \Big[ 
 \frac{ q^2 }{ 4 m_h^2 \, k^2}
\,+\, \frac{k-1}{k} \,\Big]\,
& \approx & \frac {1}{4j}  \prod_{k=2}^{j} \, \Big[ 
 \frac{(2k-1)^2}{4k(k-1)} \,\Big]
\prod_{k'=1}^{j} \, \Big[1+ \frac{ q^2 }{ m_h^2 \, (2k'-1)^2}
\,\Big]\cr
& \approx & \frac {1.27}{4j} \, \cosh\left(\frac{ \pi q }{2 m_h}\right)\,,
\label{pochhammer4}
\end{eqnarray}

\noindent which is valid since $j \ge n$ is very large. Substituting this result in eq. (\ref{Integralbrane2}), one finds 

\begin{eqnarray}
{\cal I}_{brane} & \approx & (-1)^n \frac {1.27}{4} \Gamma (\ell + 2) \Gamma (\ell + 3)
\frac{ \Gamma (n -1) } {\Gamma (n+2\ell + 5)}  \cosh\left(\frac{ \pi q }{2 m_h}\right) \,\,,
\label{Integralbrane3}
\end{eqnarray}

\noindent and then the structure function reads 

\begin{equation}
{{F_2}\vert}_{x<<1} \approx \pi^5 {\cal Q}^2  \, \frac {(1.27)^2}{2}\, \Gamma(2\ell + 4)\, (\ell+2)^2 
\left(\frac{ 4 m_h^2  }{q^2}\right)^{\ell+2}\, x^{\ell+4} \,.
\end{equation}

\noindent {\bf Elastic form factor} 

We have to work out the sum in eq. (\ref{Integralbrane2})
for the case $n=0$:
\begin{eqnarray}
{\cal I}^{^{(n=0)}}_{brane} &=& \,\frac{\Gamma (\ell + 2)}{\Gamma(2\ell+4)} \, \sum_{j=0}^\infty \,\frac{ ( -\alpha)_j (1+\alpha)_j (j+2)_{\ell}
  }{ j!  (2 \ell + 4)_j} \,\,\nonumber \\
&=& \frac{\Gamma (\ell + 2)}{\Gamma(2\ell +4)} \, \Big\{\frac{d^\ell}{dt^\ell}[F(-\alpha,1+\alpha; 2\ell +4 ; t) \, t^{\ell+1}]\Big\}_{t=1} \, .
\end{eqnarray}

Using the property 
\begin{equation}
\frac{d^\ell}{dt^\ell}F(a,b;c;t)=\frac{(a)_\ell(b)_\ell}{(c)_\ell}F(a+\ell,b+\ell;c+\ell;t) , 
\end{equation}

\noindent we find 
\begin{eqnarray}
{\cal I}^{^{(n=0)}}_{brane}&=& \frac{\Gamma^2(\ell + 2)\Gamma(\ell+1)}{\Gamma(2\ell+4)} \sum_{k=0}^\ell \Big[ \frac{1}{k! \, \Gamma(\ell+1-k)\, \Gamma(\ell+2-k)}  \frac{(-\alpha)_{\ell-k}(1+\alpha)_{\ell-k}}{(2\ell+4)_{\ell-k}} \nonumber \\
&\times&F(-\alpha+\ell-k,1+\alpha+\ell-k \, ; 3\ell+4-k \, ;1)\Big] \nonumber \\
&=& \frac{\Gamma^2(\ell + 2)\Gamma(\ell+1)}{\Gamma(-\alpha)\Gamma(1+\alpha)} \sum_{k=0}^\ell  \frac{\Gamma(\ell+3+k)}{k! \, \Gamma(\ell+1-k) \Gamma(\ell+2-k)}\Big[\prod_{k'=\ell+1-k}^{2\ell+3}[\frac{q^2}{4m_h^2}+k'(k'-1)] \Big]^{-1}. \nonumber \\
\end{eqnarray}

The relevant term in the sum, for $q^2>>m_h^2$, is $k=0$. Using eq. (\ref{branematrix}) we find 

\begin{eqnarray}
F(q^2)&=& \frac{Q}{2} \,  \Gamma(2 \ell + 4)(\ell+2)\Big(\, \frac{4m_h^2}{q^2}\Big)^{\ell+2} \nonumber \\
&=& \frac{Q}{2}\, \Gamma(2\Delta-2)(\Delta-1)\Big(\, \frac{4m_h^2}{q^2}\Big)^{\Delta-1} \, .
\end{eqnarray}

 \end{document}